\documentclass[conference]{IEEEtran}
\IEEEoverridecommandlockouts

\usepackage{graphicx}
\usepackage{xcolor}

\usepackage{amsmath}
\usepackage{amssymb}
\usepackage{cite}
\usepackage{cuted}
\newcommand{\linebreakand}{%
  \end{@IEEEauthorhalign}
  \hfill\mbox{}\par
  \mbox{}\hfill\begin{@IEEEauthorhalign}
}    
\usepackage[colorlinks=Flase]{hyperref}
\usepackage{bbm}
\usepackage{algorithm} 
\usepackage{algpseudocode}
\usepackage{caption}
\usepackage{steinmetz}
\usepackage{subcaption}
\usepackage{amsthm}
\usepackage[english]{babel}


\newcounter{relctr} 
\everydisplay\expandafter{\the\everydisplay\setcounter{relctr}{0}} 
\DeclareMathOperator*{\argmax}{argmax} 
\DeclareMathOperator*{\maximize}{maximize} 
\usepackage[figurename=Fig.]{caption}
\newcommand{\norm}[1]{\left\lVert#1\right\rVert}
\usepackage{xpatch}

\newtheoremstyle{remarkstyle}%
  {}
  {}
  {\itshape}
  {}
  {\itshape}
  {.}
  {.5em}
  {}

\theoremstyle{remarkstyle}

\AtBeginDocument{} 

\ifCLASSINFOpdf
\else
\fi

\begin{document}

\title{Sense-then-Charge: Wireless Power Transfer to Unresponsive Devices with Unknown Location\\
\thanks{This work is partially supported in Finland by the Research Council of Finland (Grants 348515, 362782, and 369116 (6G Flagship)); by the European Commission through the Horizon Europe/JU SNS projects Hexa-X-II (Grant no. 101095759) and AMBIENT-6G (Grant 101192113); by the Finnish-American Research \& Innovation Accelerator; in Brazil by CNPq (305021/2021-4) and RNP/MCTI (01245.020548/2021-07) 6G Mobile Communications Systems; and in Denmark by the Villum Investigator Grant “WATER” from the Velux Foundation.}
}

\author{
{ Amirhossein~Azarbahram$^{*}$, Onel~L.~A.~L\'{o}pez$^{*}$, Richard~D.~Souza$^{\circ}$, Petar~Popovski$^{\dagger}$, Matti~Latva-aho$^{*}$}
\vspace{2mm}
\\

\small	$^{*}$Centre for Wireless Communications (CWC), University of Oulu, Finland\\
$^{\circ}$Department of Electrical and Electronics
Engineering, Federal University of Santa Catarina, Florianopolis,
Brazil \\
$^{\dagger}$Department of Electronic Systems, Aalborg University, Denmark \\
\small Emails: \{amirhossein.azarbahram, onel.alcarazlopez\}@oulu.fi, richard.demo@ufsc.br, petarp@es.aau.dk, matti.latva-aho@oulu.fi}


\maketitle

\begin{abstract}


This paper explores a multi-antenna dual-functional radio frequency (RF) wireless power transfer (WPT) and radar system to charge multiple unresponsive devices. We formulate a beamforming problem to maximize the minimum received power at the devices without prior location and channel state information (CSI) knowledge. We propose dividing transmission blocks into sensing and charging phases. First, the location of the devices is estimated by sending sensing signals and performing multiple signal classification and least square estimation on the received echo. Then, the estimations are used for CSI prediction and  RF-WPT beamforming. Simulation results reveal that there is an optimal number of blocks allocated for sensing and charging depending on the system setup. Our sense-then-charge (STC) protocol can outperform CSI-free benchmarks and achieve near-optimal performance with a sufficient number of receive antennas and transmit power. However, STC struggles if using insufficient antennas or power as device numbers grow.

\end{abstract}


\IEEEpeerreviewmaketitle

\section{Introduction}

\IEEEPARstart{R}{adio} frequency (RF) wireless power transfer (WPT) is a promising technology for supporting uninterrupted connectivity among devices in the future Internet of Things (IoT). This can be done by providing multi-user wireless charging using existing wireless communications infrastructure \cite{ZEDHEXA, 3gppamb, lópez2023highpower}. However, the efficiency of RF-WPT, referred to simply as WPT in this work, remains a key challenge, requiring further optimization research. Energy beamforming (EB) is a promising solution for compensating for the channel losses by directing the transmit signal toward the receivers. However, most EB-related works, e.g., \cite{SISOAllClreckx,onellowcomp, MyEBDMA}, rely on available channel state information (CSI) or perfect knowledge of the location of devices to optimize system performance, while a few others focus on the CSI acquisition, e.g., \cite{WPT-up_training, OnelRadioStripes}, or EB with imperfect CSI, e.g., \cite{statistical_imperfect_wet}. 

\begin{figure}
    \centering
    \includegraphics[width=\columnwidth]{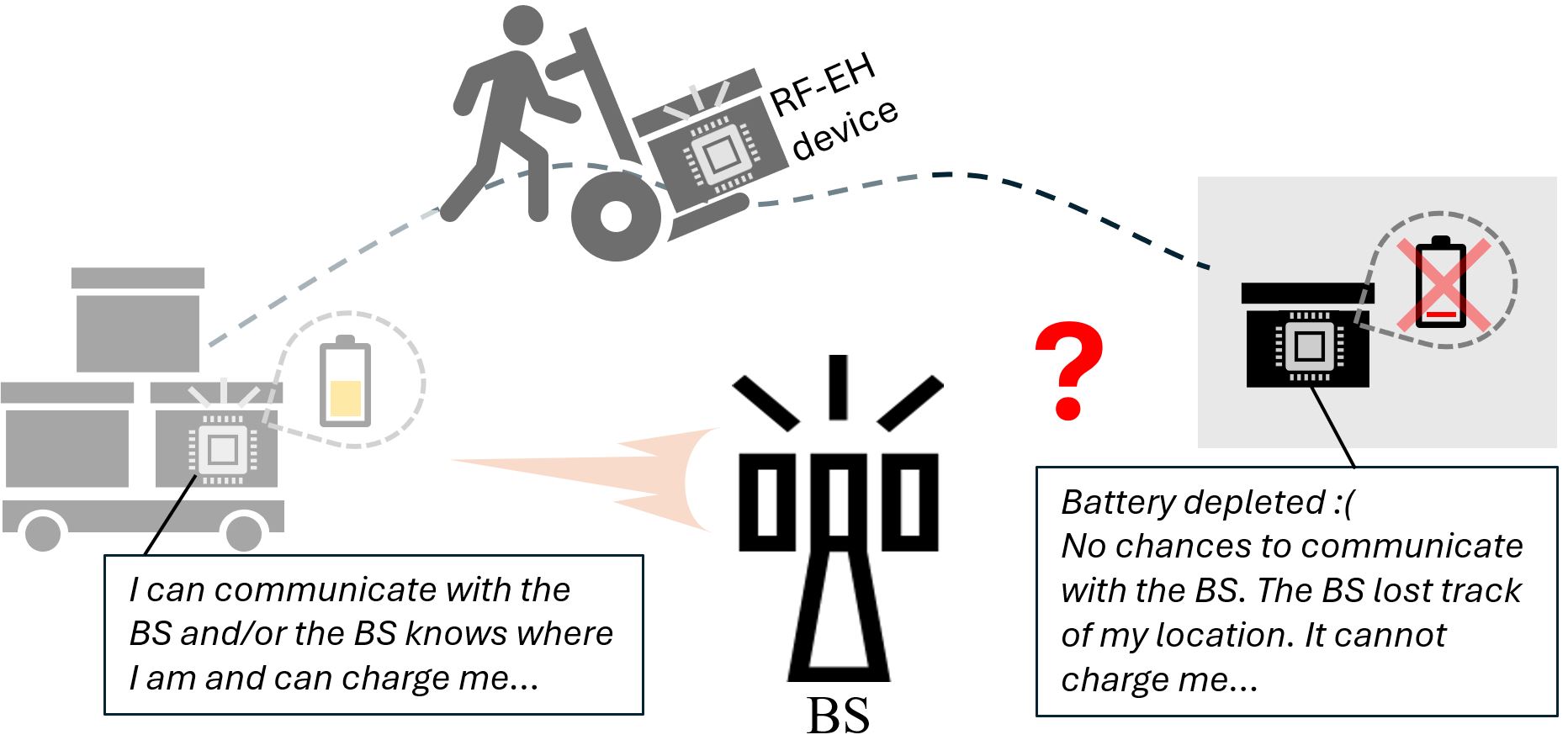}
    \caption{An example scenario where CSI or location is unavailable and the nodes are unresponsive, where IoT nodes attached to a moving object overspent their energy.}
    \label{fig:usecase}
\end{figure}

In wireless systems where devices are unresponsive due to, e.g., battery depletion, and cannot communicate with the charging base station (BS), the lack of CSI makes EB particularly challenging. To cope with this, one can rely on the location information of the devices, as in \cite{loc-based-access}, where a novel form of beam diversity is exploited to increase the initial access range of the devices. However, in some cases, no prior knowledge of the location of the devices is available due to, e.g., their movement after becoming unresponsive. A highly relevant scenario arises when the devices are attached to moving objects and their location changes after becoming unresponsive as in Fig.~\ref{fig:usecase}. Another example is related to disaster-stricken areas \cite{disaster_paper}, where critical devices, such as sensors providing valuable information about the environment are nonfunctional while their position is unknown. In such cases, BS needs to wirelessly recharge these devices without any signaling and prior precise location assumption such that they become operational and resume vital communications. Thus, the devices must first receive sufficient energy above a sensitivity threshold, which is the minimum energy required for wake-up or backscattering \cite{wet_feas_mMIMO}, letting them start communicating with the BS and obtaining CSI afterward. Again, CSI-based EB is not an option since conventional CSI acquisition, relying on pilot transmission, is not possible due to the unresponsive state of the device. Instead, CSI-free multi-antenna EB strategies can be leveraged to charge the devices, as in \cite{onel_csifree, onel_rotary}, which focus on solutions for massive IoT deployments. In \cite{onel_rotary}, a rotary antenna beamforming (RAB) approach is proposed to charge the devices without relying on CSI, which outperforms the CSI-free approaches proposed in \cite{onel_csifree}. Notably, CSI-free EB becomes highly inefficient as the number of IoT devices to be charged decreases since the RF energy lost in undesired spatial directions increases \cite{onel_csifree, onel_rotary}.


To overcome the challenges of CSI-free EB, integrated sensing capabilities of future wireless systems may come to help by enabling the networks to measure and even image their surroundings \cite{fanliu_isac_survey, crb_opt_fanliu, ISAC_multitarget, OPT_ISAC_noise_ref, RIS_CRB_ISAC}. In this work, we investigate one of the potential roles of sensing for CSI-free WPT by considering the initial access phase of a multi-user dual-functional system, capable of both WPT and sensing. We consider that the devices are unresponsive and the BS has no prior knowledge of their location. We propose a sense-then-charge (STC) protocol, which divides the system's operation time between sensing and charging. First, the BS sends sensing signals and estimates the line of sight (LoS) channel of the energy harvesting (EH) devices based on the received echo signal and by relying on the multiple signal classification (MUSIC) \cite{musicref} and least square estimation. Then, the BS charges the devices using the estimated LoS channel. This is fundamentally different from the other works focusing on integrated sensing and WPT, e.g., \cite{ISWPT, isac+swipt}. Specifically, \cite{ISWPT} considers separate EH receivers and radar targets, while in \cite{isac+swipt}, the BS has prior information on the angle of arrival (AoA) and reflection coefficient of the devices, which corresponds to the tracking phase of a slowly moving target \cite{crb_opt_fanliu}. Our simulation results demonstrate that the proposed STC protocol can outperform the CSI-free benchmarks in \cite{onel_rotary, onel_csifree} and achieve near-optimal performance given a sufficient number of receive antennas and transmit power. Moreover, we shed some light on the charging and sensing performance trade-offs depending on the number of transmit and receive antennas and the time allocated to each phase.

\textbf{{Structure:}} Section~\ref{sec:system1} presents the WPT system model and problem formulation. Section~\ref{sec:system2} illustrates the proposed sensing-based charging mechanism along with sensing and WPT beamforming approaches. Section~\ref{sec:perfromancemetric} introduces the performance metrics, Section~\ref{sec:numerical} provides the numerical analysis, and Section~\ref{sec:conclude} concludes the paper.

\textbf{{Notations:}} Bold lower-case and upper-case letters represent vectors and matrices, respectively. The $\ell_2$-norm operator is denoted by $\norm{\cdot}$. Moreover, $(\cdot)^H$ and $(\cdot)^\star$ denote the transposed conjugate and conjugate operations, respectively. The vectorization operator is represented by $\mathrm{Vec}(\cdot)$, $\mathbf{I}_D$ represents a $D \times D$ identity matrix, and $\mathrm{diag}(\mathbf{a})$ refers to a diagonal matrix with its main diagonal being the elements of vector $\mathbf{a}$.

{\textbf{Reproducible research:}} Simulation codes are available at: \url{https://github.com/amiraz96/STC-WPT}

\section{WPT System Model \& Problem Formulation}\label{sec:system1}

We consider a BS equipped with a uniform linear array (ULA) and $N$ antennas. Moreover, $K$ unresponsive single-antenna EH devices are located within the coverage area of the BS. The BS aims to charge the devices without any prior knowledge of their location over $L$ transmission blocks, although the value $K$ is assumed known for simplicity. The transmit power budget per block is considered to be $P_t$. 

We consider far-field WPT such that $\beta_k\mathbf{h}_k(\theta_k) \in \mathbb{C}^{N \times 1}$ is the downlink (DL) channel of the $k$th device, where $\theta_k$ is the angle of departure, $\beta_k$ is the path gain, and $\mathbf{h}_k(\theta_k)$ is the small-scale fading between the BS and $k$th device. We adopt a quasi-static Rician fading model such that 
\begin{equation}
    {\mathbf{h}}_k(\theta_k) = \sqrt{\kappa/(\kappa + 1)}\mathbf{a}(\theta_k) + \sqrt{1/(\kappa + 1)}\mathbf{b}_k,
\end{equation}
where $\kappa$~dB is the Rician factor, $\mathbf{a}({\theta})$ is the $N$-dimensional steering vector, and $\mathbf{b}_k$ is the non-LoS (NLoS) component of the channel.
For simplicity, we assume that the channel is static during $L$ blocks. The received charging signal at the $k$th device and $l$th block is given by
\begin{equation}
    \mathbf{y}_{k}[l] = \sum_{i = 1}^{K} {\beta_k} 
 \mathbf{h}_k(\theta_k)^H \mathbf{w}_{i} {s}_{i}[l],
\end{equation}
where ${s}_{i}[l]$ denotes the $i$th energy symbol at the $l$th block and $\mathbf{w}_{i} \in \mathbb{C}^{N \times 1}$ is the corresponding transmit beamforming vector. Moreover, the impact of noise is not considered, as it is typically negligible in WPT systems \cite{OnelRadioStripes}. We consider normalized independent symbols such that $\mathbb{E}_s\bigl\{{s}_{i}[l]{{s}_{i}[l]}^\star \bigr\} = 1$ and $\mathbb{E}_s\bigl\{{s}_{i}[l]{{s}_{r}[l]}^\star \bigr\} = 0$. Hereby, the received power by the $k$th device is given by
\begin{equation}\label{eq:recvpower}
    \tilde{P}_{k} = \mathbb{E}_s\big\{\mathbf{y}_{k}[l] {\mathbf{y}_{k}[l]}^H\big\} = {\sum_{i = 1}^{K} \norm{{\beta_k} 
 \mathbf{h}_k(\theta_k)^H\mathbf{w}_{i}}^2}.
\end{equation}

The goal is to maximize the minimum RF power at the devices given a transmit power budget to help them reach a sufficient amount of energy for regaining functionality.  Thus, the problem can be written as
\begin{subequations}\label{main_prob}
\begin{align}
\label{main_proba} \maximize_{\mathbf{w}_{i}} \quad &  \min_k \mathbb{E}_{\beta_k, \theta_k}\bigg\{{\sum_{i = 1}^{K} \norm{{\beta_k} 
 \mathbf{h}_k(\theta_k)^H \mathbf{w}_{i}}^2}\bigg\}\\
\textrm{subject to} \label{main_probc}  \quad & \sum_{k = 1}^K \norm{\mathbf{w}_{k}}^2 \leq P_t,
\end{align}
\end{subequations}
where the expectation in the objective function is w.r.t. $\beta_k, \theta_k$ since these are the unknown parameters of the system depending on the deployment. Notably, this problem is addressed in \cite{onel_csifree, onel_rotary}, but only for uniform $\theta \in [0, 2\pi]$ and $\beta_k$ independent of $\theta_k$. Thus, the solutions proposed therein are only applicable to large uniform IoT deployments.

In general, problem \eqref{main_prob} is impossible to solve optimally since the BS has no prior knowledge about the location of the EH devices or CSI, leading to completely unknown $\mathbf{h}_k, \forall k$. Thus, we propose the STC protocol to estimate the location of the devices and perform WPT beamforming relying on the estimated LoS channel.

\section{STC Protocol}\label{sec:system2}

\begin{figure}
    \centering
    \includegraphics[width=0.85\columnwidth]{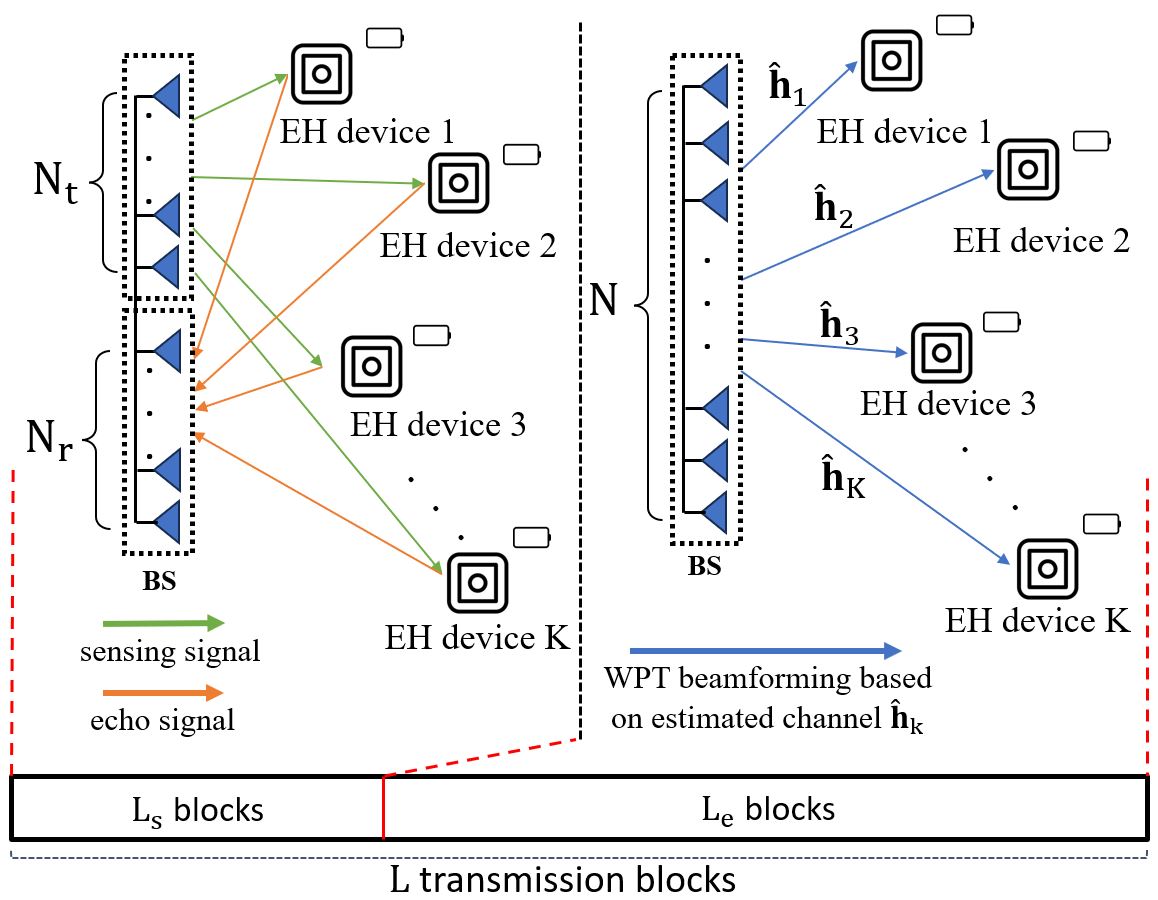}
    \caption{Dual-functional sensing-enabled WPT system. The devices are treated as sensing targets during the first $L_s$ blocks. Then, they are charged during the rest of the blocks by using the estimated LoS channel in the sensing phase.}
    \label{fig:swpt}
\end{figure}

Let us divide the transmission blocks such that the devices are treated as sensing targets during the first $L_s$ blocks, aiming to estimate their location. Then, the remaining $L_e = L - L_s$ blocks are used to charge the targets based on the estimated device locations. This division introduces a trade-off between the location estimation accuracy in the sensing phase and the WPT efficiency in the charging phase, suggesting the existence of an optimum configuration $(L_s, L_e)$ given a fixed $L$ as corroborated in Section~\ref{sec:numerical}. Moreover, note that the BS needs precise and directed beams to charge the devices closer to the coverage area edge, requiring larger $L_s$ and $L_e$ (and $L$) for reliable sensing and sufficient energy harvesting, respectively. We consider that $N_r > K$ and $N_t$ antennas are utilized for reception and transmission at the BS, respectively, such that $N_t + N_r = N$. Thus, the system operates in two phases, namely, i) target sensing with $N_t$ and $N_r$ transmit and receive antennas, and ii) target charging with all $N$ antennas. The sensing-enabled WPT system is illustrated in Fig.~\ref{fig:swpt}.

\subsection{Sensing Signal Transmission and Reception}\label{subsec:sensing_model}

We consider that the BS sends a sensing signal over $L_s$ sensing blocks formulated as $\mathbf{X} \in \mathbb{C}^{N_t \times L_s} = \mathbf{\bar{W}} \mathbf{\bar{S}}$, where $\mathbf{\bar{W}} \in \mathbb{C}^{N_t \times N_t} = [\mathbf{\bar{w}}_{1}, \ldots, \mathbf{\bar{w}}_{N_t}]$ is the sensing beamforming matrix and $\mathbf{\bar{S}} \in \mathbb{C}^{N_t \times L_s}$ is the matrix containing the unit-power independent sensing symbols such that $\mathbf{\bar{S}}\mathbf{\bar{S}}^H/L_s = \mathbf{I}_{N_t}$. Notably, we consider $N_t$ individual streams for sensing to leverage all spatial degrees of freedom for probing the targets \cite{fanliu_isac_survey}. Then, the received echo signal is given by\footnote{We ignore self-interference as previous works on radar sensing \cite{isac+swipt, crb_opt_fanliu} since the sensing receiver can handle it properly.}
\begin{equation}\label{eq:echosignal}
    \mathbf{Y} = \mathbf{G}\mathbf{X} + \mathbf{Z},
\end{equation}
where $\mathbf{G} \in \mathbb{C}^{N_r \times N_t}$ is the response matrix of the targets depending on their round-trip path loss, reflection characteristics, angular direction, and other environmental factors such as clutter. Moreover, $\mathbf{Z} \in \mathbb{C}^{N_r \times L_s}$ is the zero-mean additive white Gaussian noise (AWGN) with a variance $\sigma^2$. 

Let us proceed by assuming that i) each target is a point target sufficiently far from the BS such that the radar is mono-static, leading to the same direction of arrival and departure \cite{fanliu_isac_survey}; ii) the targets are located within the same range resolution of the radar, as in \cite{ISAC_multitarget}, allowing for $\mathbf{G}$ to be written as the superposition of the responses of the targets; iii) the reflectivity of clutters in the environment is sufficiently weak compared to the sensing targets.\footnote{This implies that the targets exhibit stronger characteristic reflections due to higher reflectivity or favorable positioning, allowing them to be reliably detected and separated from background clutter.  This can be facilitated with special reflective materials and manufacturing techniques \cite{lopez2024leakywaveantennaequippedrf}.} Then, by defining $\mathbf{h}_{k, t}(\theta_k) \in \mathbb{C}^{N_t \times 1}$ and $\mathbf{h}_{k, r}^T(\theta_k) \in \mathbb{C}^{N_r \times 1}$ as the transmit and receive channels, the reflection matrix is given by \cite{ISAC_multitarget, crb_opt_fanliu}
\begin{equation}\label{eq:reflkectionmat}
    \mathbf{G} = \sum_{k = 1}^K \alpha_k \mathbf{h}_{k, r}(\theta_k)\mathbf{h}_{k, t}^T(\theta_k),
\end{equation}
where $\alpha_k$ includes the round-trip path loss and object reflection properties. Specifically, we can write $\alpha_k = \rho_k {\beta_k}^2$, where $\rho_k$ is radar cross section (RCS) \cite{fanliu_isac_survey} of the $k$th target.

\subsection{Sensing Beamforming and Parameter Estimation}

Since there is no prior knowledge of target parameters, i.e., $\theta_k, \alpha_k$, we need to estimate the whole response matrix. Thus, we consider the CRB for $\mathbf{G}$ given by \cite{crb_opt_fanliu, statics_ig_book, fanliu_isac_survey} 
\begin{equation}
    {\mathrm{CRB}}(\mathbf{\bar{W}}) = ({\sigma^2  N_r}/{L_s})\mathrm{Tr}(\mathbf{R}_x^{-1}),
\end{equation}
where $\mathbf{R}_x = \frac{1}{L_s}\mathbf{X}\mathbf{X}^H = \frac{1}{L_s}\mathbf{\bar{W}} \mathbf{S}_s \mathbf{S}_s^H \mathbf{\bar{W}}^H = \mathbf{\bar{W}} \mathbf{\bar{W}}^H$ is the sample covariance matrix of $\mathbf{X}$. Hereby, the sensing beamforming design problem aims to minimize ${\mathrm{CRB}}(\mathbf{\bar{W}})$ such that $\norm{\mathbf{\bar{W}}}_F^2 \leq  P_t$, which is proven to be optimally solved by setting $\mathbf{R}_x = ({ P_t}/{N_t})\mathbf{I}_{N_t}$ \cite{crb_opt_fanliu}, leading to $\bar{\mathbf{W}} = \sqrt{P_t/N_t}\mathbf{I}_{N_t}$.


Now, we aim to estimate the location parameters of the targets, i.e., angle and path gain, using the received echo signal at the BS to leverage this information later for EH purposes. For this, we use the well-known MUSIC algorithm \cite{musicref} to estimate the AoA of the targets. 

Let $\mathbf{R}_y = \frac{1}{L_s}\mathbf{Y}\mathbf{Y}^H$ be the sample covariance matrix of the received echo signal, which can be written using the eigenvalue decomposition (EVD) as $\mathbf{R}_y = \mathbf{U \Lambda U}^H$. Moreover, $\mathbf{U} = [\mathbf{U}_s, \mathbf{U}_n]$ consists of the signal and noise subspace and $\mathbf{\Lambda} = \mathrm{diag}[\lambda_1, \ldots, \lambda_K, \sigma^2, \ldots, \sigma^2]^T$, where $\lambda_k$ is the $k$th largest eigenvalue. Since the noise subspace is orthogonal to the steering vectors, the MUSIC spectrum is given by \cite{musicref}
\begin{equation}\label{eq:MUSIC_SPECTRUM}
    P_{\text{MUSIC}}(\theta) = \frac{1}{\mathbf{a}_r^H(\theta) \mathbf{U}_n \mathbf{U}_n^H \mathbf{a}_r(\theta)},
\end{equation}
where $\mathbf{a}_r(\theta) \in \mathbb{C}^{N_r \times 1}$ is the receive steering vector for $\theta$ and the estimated AoAs are the $K$ largest peaks of \eqref{eq:MUSIC_SPECTRUM}.

Let $\mathbf{C} = \mathrm{diag}([\alpha_1, \ldots, \alpha_{K}]^T)$, then, the echo signal can be rewritten as 
\begin{equation}
    \mathbf{Y} = \mathbf{H}_r(\boldsymbol{\theta})\mathbf{C}\mathbf{H}_t^T(\boldsymbol{\theta})\mathbf{X} + \mathbf{Z},
\end{equation}
where $\mathbf{H}_t(\boldsymbol{\theta}) = [\mathbf{h}_{1, t}(\theta_1), \ldots, \mathbf{h}_{K, t}(\theta_K)]^T$ and $\mathbf{H}_r(\boldsymbol{\theta}) = [\mathbf{h}_{1, r}(\theta_1), \ldots, \mathbf{h}_{K, r}(\theta_K)]^T$. By defining $\mathbf{B} = \mathbf{H}_t^T(\boldsymbol{\theta})\mathbf{X}$ and $\bar{\mathbf{B}} = \mathbf{B}^T \otimes \mathbf{H}_r(\boldsymbol{\theta})$,  we can write 
\begin{equation}
    \mathrm{Vec}(\mathbf{Y_s}) = \bar{\mathbf{B}} \mathrm{Vec}(\mathbf{C}) + \mathrm{Vec}(\mathbf{Z}).
\end{equation}
Although the channel is not necessarily LoS, we rely on the estimated LoS component to perform the coefficient estimation since we do not have any knowledge of the NLoS components. Thus, we consider $\hat{\mathbf{h}}_{k, t}(\theta_k) = \mathbf{a}_t(\hat{\theta}_k)$ and $\hat{\mathbf{h}}_{k, r}(\theta_k) = \mathbf{a}_r(\hat{\theta}_k)$, where $\mathbf{a}_t(\theta)$ is the $N_t$-dimensional transmit steering vector. Hereby, the least square estimation\footnote{Note that since the system deals with zero-mean AWGN with constant variance, least square is equivalent to maximum likelihood estimation.} of the reflection coefficients is given by
\begin{equation}\label{eq:vecC}
    \mathrm{Vec}(\hat{\mathbf{C}}) = (\bar{\mathbf{B}}^H \bar{\mathbf{B}})^{-1}\bar{\mathbf{B}}^H\mathrm{Vec}(\mathbf{Y_s}),
\end{equation}
where the estimated $\hat{{\alpha}}_k$ is the $k$th diagonal entry of $\hat{\mathbf{C}}$.

\subsection{WPT Beamforming}

WPT EB is performed by using all $N$ antennas, relying on LoS channel estimation, and ignoring potential small-scale fading. Note that  $\hat{{\alpha}}_k$ captures both $\beta_k$ and the target reflection properties.  Hereby and by denoting $\hat{\beta}_k = f(\hat{\alpha}_k)$, the WPT beamforming problem is rewritten as 
\begin{subequations}\label{WPTBF}
\begin{align}
\label{WPTBFa} \maximize_{\mathbf{w}_{i}} \quad &  \min_k \sum_{i =1}^K \norm{{f(\hat{\alpha}_k)} \mathbf{a}(\hat{\theta}_k)^H \mathbf{w}_{i}}^2\\
\textrm{subject to} \label{WPTBFb}  \quad & \sum_{k = 1}^K \norm{\mathbf{w}_{k}}^2 \leq P_t.
\end{align}
\end{subequations}
By assuming that all the targets have the same reflection properties\footnote{This assumption is valid in real-world applications, particularly when the targets are made of similar materials with uniform electromagnetic properties.} such that $\rho_k = \rho$, we can write  $f(\hat{\alpha}_k) \propto \sqrt{|\hat{\alpha}_k|}$. Thus, \eqref{WPTBFa} can be equivalently written as $\min_k \sum_{i =1}^K {|\hat{\alpha}_k|} \norm{\mathbf{a}^H(\hat{\theta}_k) \mathbf{w}_{i}}^2$. The resulting problem is not immediately convex, but it can be rewritten as \cite{onellowcomp}
\begin{subequations}\label{WPTBFreform}
\begin{align}
\label{WPTBFreforma} \maximize_{t,\ \mathbf{W} \succeq 0} \quad &  t \\
\textrm{subject to} \label{WPTBFreformb} \quad & t \leq \mathrm{Tr}(\mathbf{W}\hat{\mathbf{H}}_k), \forall k,\\
\label{WPTBFreformc}  \quad & \mathrm{Tr}(\mathbf{W}) \leq P_t,
\end{align}
\end{subequations}
where $\mathbf{W} = \sum_{i = 1}^{K} \mathbf{w}_{i}\mathbf{w}_{i}^H$ and $\hat{\mathbf{H}}_k = |\hat{\alpha}_k| \mathbf{a}(\hat{\theta}_k)\mathbf{a}^H(\hat{\theta}_k)$. This is now a semi-definite program, which can be solved by standard convex optimization tools such as CVX \cite{cvxref}. Then, $\mathbf{w}_{i}, \forall i$ can be obtained as the eigenvectors of $\mathbf{W}$ multiplied by their corresponding singular value.

\begin{algorithm}[t]
	\caption{STC protocol for multi-user WPT.} \label{alg:senseWPT}
	\begin{algorithmic}[1]
            \State \textbf{Input:} $L_e, L_s, N_t, N_r$ 
            \State \textbf{Output:} $\hat{\theta}_k, \hat{\alpha}_k, \forall k$, $\mathbf{w}^*_{i}, \forall i$
            \State Design the sensing beamformer $\bar{\mathbf{W}} = \sqrt{P_t/N_t}\mathbf{I}_{N_t}$
            \State Design the sensing symbols such that $\mathbf{\bar{S}}\mathbf{\bar{S}}^H/L_s = \mathbf{I}_{N_t}$
            \State Send $\mathbf{X} = \mathbf{\bar{W}} \mathbf{\bar{S}}$ and collect the received echo signal $\mathbf{Y}$ 
            \State Compute the sampled covariance matrix $\mathbf{R}_y = \mathbf{Y}\mathbf{Y}^H/L_s$\label{music_start}
            \State  Write the EVD of $\mathbf{R}_y$ as $\mathbf{R}_y = \mathbf{U \Lambda U}^H$
            \State Compute the signal and noise subspaces $\mathbf{U} = [\mathbf{U}_s, \mathbf{U}_n]$
            \State Compute \eqref{eq:MUSIC_SPECTRUM} for $\theta \in [-\pi/2, \pi/2]$
            \State Choose the $K$ dominant peaks of \eqref{eq:MUSIC_SPECTRUM} as $\hat{\theta}_1, \ldots, \hat{\theta}_K$ \label{music_end}
            \State Set $\hat{\mathbf{h}}_{k, t}(\theta_k) = \mathbf{a}_t(\hat{\theta}_k), \quad \hat{\mathbf{h}}_{k, r}(\theta_k) = \mathbf{a}_r(\hat{\theta}_k),\quad \forall k$
            \State Compute $\mathbf{B} = \hat{\mathbf{H}}_t^T(\hat{\boldsymbol{\theta}})\mathbf{X}$ and $\bar{\mathbf{B}} = \mathbf{B}^T \otimes \hat{\mathbf{H}}_r(\hat{\boldsymbol{\theta}})$
            \State Compute $\mathrm{Vec}(\hat{\mathbf{C}})$ using \eqref{eq:vecC}
            \State Select the diagonal entries of $\hat{\mathbf{C}}$ as $\hat{\alpha_k}, \forall k$ 
            \State Solve \eqref{WPTBFreform} for $\hat{\mathbf{H}}_k = |\hat{\alpha}_k| \mathbf{a}(\hat{\theta}_k)\mathbf{a}^H(\hat{\theta}_k), \forall k$ to obtain $\mathbf{W}^*$
            \State Obtain $\mathbf{w}^*_{i}, \forall i$ as the eigenvectors of $\mathbf{W}^*$ multiplied by the corresponding singular value
            \State Send $K$ charging signals at each of $L_e$ WPT blocks such that $\mathbf{w}^*_{i} {s}_{i}[l]$ is the $i$th signal at the $l$th block.
\end{algorithmic} 
\end{algorithm}

Algorithm~\ref{alg:senseWPT} illustrates the proposed STC protocol. First, the sensing signal $\mathbf{X}$ is designed and sent by the transmit antennas, and the echo signal $\mathbf{Y}$ is collected at the receive antennas. Then, the MUSIC algorithm is used in lines \ref{music_start}-\ref{music_end} to estimate $\theta_k$ values. After that, $\alpha_k$ values are estimated using least square estimation. Then, WPT beamforming vectors are designed by relying on the estimated LoS channel. Finally, the devices are charged by the designed charging signals using the obtained WPT beamforming vectors.

\section{Performance Metrics}\label{sec:perfromancemetric}

Herein, we present the performance metrics to evaluate the STC protocol. For the accuracy of the estimation, we use the relative error computed by $\sum_{k = 1}^K (\theta_k - \hat{\theta}_k)/K$, $\sum_{k = 1}^K |\alpha_k - \hat{\alpha}_k|/(K|\alpha_k|)$, and $\sum_{k = 1}^K \norm{\beta_k \mathbf{a}(\theta_k) - \sqrt{|\hat{\alpha}|}_k\mathbf{a}(\hat{\theta_k}) }/(KN{\beta_k})$ for AoAs, $\alpha_k$ values, and LoS channel, respectively.


Note that the devices operate independently of the phase they are in and can therefore harvest energy even during the sensing blocks. This harvested energy is given by $\bar{E}_{k} \propto \bar{P}_{k} = {\sum_{i = 1}^{N_t} \norm{\beta_k \mathbf{h}(\theta_k)^H \mathbf{\bar{w}}_{i}}^2}$. Hereby, assuming equal time duration for blocks, we denote the total received energy during the WPT and sensing phases as $\tilde{E}_{k} \propto L_e \tilde{P}_{k}$ and $\bar{E}_{k} \propto L_s \bar{P}_{k}$, respectively. One can compute the harvested energy in Joule units by considering block durations and summing the received power over the operating time of the system. In this paper, without loss of generality, we consider normalized block durations and choose $P_k = L_s \bar{P}_{k} + L_e \tilde{P}_{k}$ as the RF energy impinging the $k$th device.

Note that $\tilde{P}_{k}$ and $\bar{P}_{k}$ correspond to the received RF power, which can be converted to DC by multiplying by a fixed RF-to-DC conversion efficiency when using the linear EH model. Meanwhile, proper reformulations and considering the saturation region of the EH devices are needed when considering non-linear EH, as in \cite{SISOAllClreckx, azarbahram2024waveform, azarbahram2025diagonalreconfigurableintelligentsurfaces}. 

We divide the transmission blocks using $0 \leq \gamma \leq 1$ such that $L_s = \lfloor\gamma L\rfloor$. Furthermore, each setup includes a desired $\gamma^\star  = \argmax_u \min_k P_k(\gamma_u)$ such that $\gamma = \gamma^\star$ leads to maximizing $\min_k P_k(\gamma_u)$ during $L$ blocks. For obtaining $\gamma^\star$, we perform a one-dimensional search over parameter $\gamma$ with a step size $\xi \ll 1$ such that $\gamma_u \in \{0, \xi, 2\xi, \ldots, 1\}$.

\section{Numerical Analysis}\label{sec:numerical}

We consider the inter-element distance of the ULA to be $\lambda/2$, where $\lambda$ is the associated wavelength to a WiFi-like scenario with the frequency 2.4~GHz. Hereby. the steering vector of ULA is $\mathbf{a}(\theta) = [1, e^{-j\pi \sin(\theta)}, \ldots, e^{j\pi ({N - 1})\sin(\theta)}]^T$. The large-scale fading is modeled using the free-space formulation $\beta_k = \lambda/(4\pi d_k)$, where $d_k$ is the distance between the BS and the $k$th device. 
We assume that channel reciprocity holds between the BS and the targets. Unless stated otherwise, we set $\kappa=20$~dB, which corresponds to a strong LoS influence, typical in short-range WPT setups. Furthermore, $\mathbf{b}_k$ is modeled by a zero-mean unit-variance circularly symmetric complex gaussian vector. The devices are uniformly deployed in $\theta_k \in [-80^\circ, 80^\circ]$ and $d_k \in [5 \text{m},15 \text{m}]$, while their RCS parameter is $\rho_k = \rho \sim \mathcal{CN}(0,1)$ \cite{RIS_CRB_ISAC}. The step size for obtaining $\gamma^\star$ is $\xi = 0.05$ and $\sigma^2 = -70$~dBm \cite{OPT_ISAC_noise_ref}. Finally, the results are averaged over 500 random realizations of deployments and $L = 1000$. We consider three benchmarks: \\
i) \textbf{Perfect knowledge (PK)}: This approach solves \eqref{WPTBFreform} assuming perfect $\theta_k$ and $d_k$, thus leading to an unachievable upper-bound performance, useful for benchmarking. \\ ii) \textbf{AA-IS}: This refers to all antennas sending equally powered independent symbols such that $P_k = L P_t \norm{{\mathbf{h}_k}}^2/N_t$ \cite{onel_csifree}. 
\\ iii) \textbf{RAB}: This refers to the RAB approach proposed in \cite{onel_rotary}, which leverages mechanical rotation to charge a massive number of devices without relying on CSI. However, it is important to note that this method incurs additional costs and complexity due to mechanical rotation, which are not accounted for in our analysis. Consequently, the mechanical rotation provides RAB with a distinct advantage, making a direct comparison with other approaches not entirely equitable.

\begin{figure}[t]
    \centering
    \includegraphics[width=0.5\columnwidth]{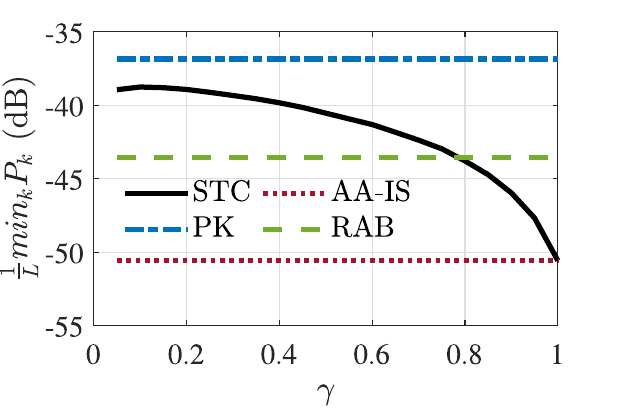} 
    \hspace{-5mm}
    \includegraphics[width=0.5\columnwidth]{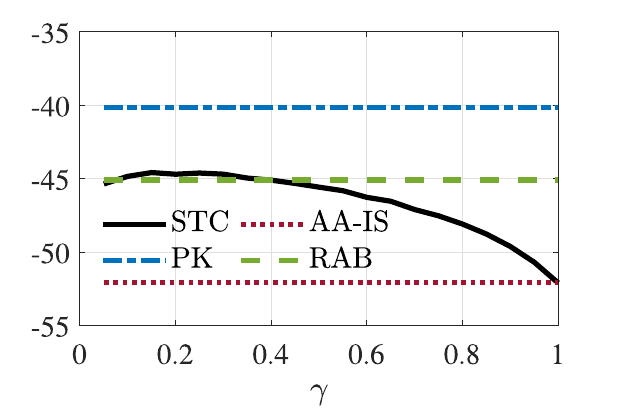} \\
    \vspace{-5.2mm}
    \includegraphics[width=0.5\columnwidth]{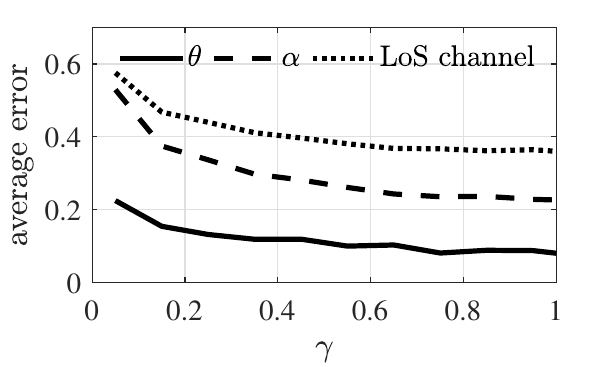} 
    \hspace{-5mm}
    \includegraphics[width=0.5\columnwidth]{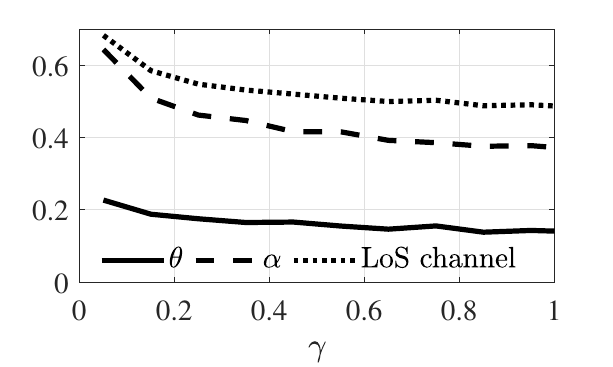}
    \caption{(a) Average minimum $P_k$ (top) and (b) average error (bottom) as a function of $\gamma$ for $K = 2$ (left) and $K = 4$ (right). We assume $N_r = 24$, $N_t = 12$, and $P_t = 10$ dB.}
    \label{fig:overgamma}
    \vspace{-0.4cm}
\end{figure}

Fig.~\ref{fig:overgamma} illustrates the average minimum $P_k$ and estimation error as a function of $\gamma$. It is observed that increasing $\gamma$ leads to lower error by allocating more blocks to sensing. However, there is a trade-off between minimum $P_k$ and error. After some point, the amount of minimum power at the devices starts to decrease. This is expected since the number of dedicated EH blocks reduces with $\gamma$ and although the estimations become more accurate, they cannot be leveraged efficiently to charge the devices. Thus, there is an optimal $\gamma$ for each setup such that the minimum $P_k$ is maximized. Moreover, it is seen that as $K$ increases, the minimum $P_k$ decreases, which is expected since acquiring proper estimations becomes more complicated. Interestingly, it is seen that STC outperforms RAB for some $\gamma$ values and it even approaches near-optimal performance for $K = 2$. Since AA-IS is a special case of STC with $\gamma = 1$, i.e., harvesting energy only from the sensing signal, it is evident that STC always outperforms AA-IS.

\begin{figure}[t]
    \centering
    \includegraphics[width=0.5\columnwidth]{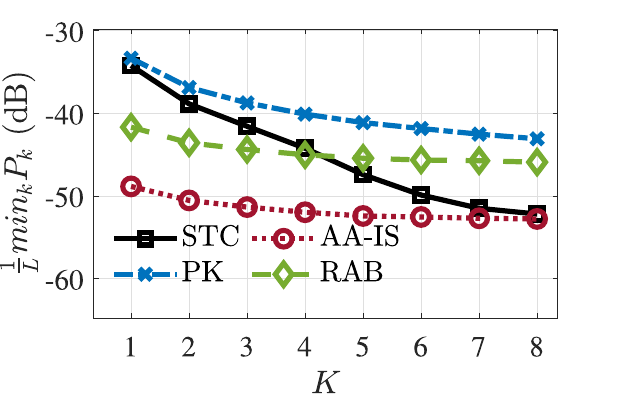} 
    \hspace{-5mm}
    \includegraphics[width=0.5\columnwidth]{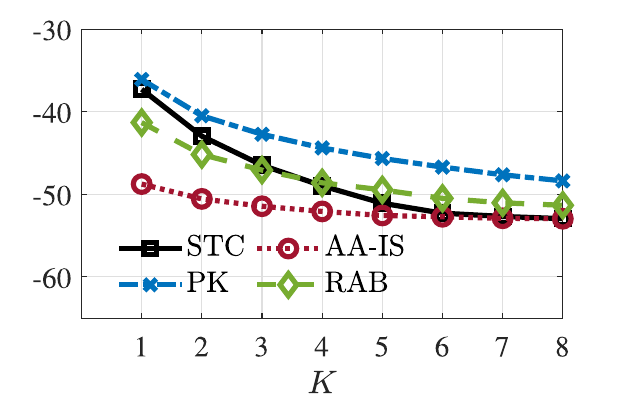} \\
    \vspace{-5.5mm}
    \includegraphics[width=0.5\columnwidth]{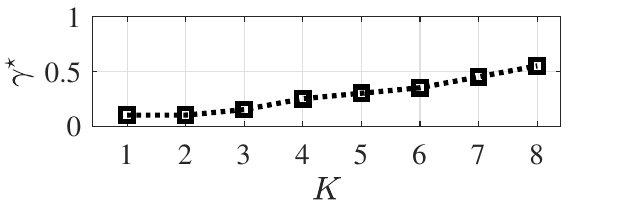} 
    \hspace{-5mm}
    \includegraphics[width=0.5\columnwidth]{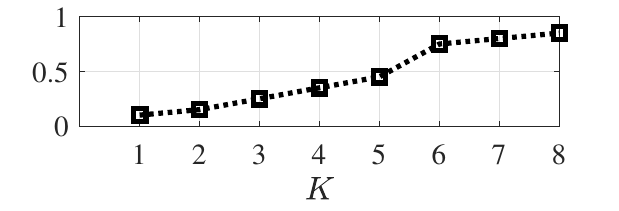}
    \caption{(a) Average minimum $P_k$ (top) and (b) $\gamma^\star$ (bottom) as a function of $K$ for $\kappa = 20$ dB (left) and $\kappa = 0$ dB (right). We assume $N_r = 24$ and $N_t = 12$.}
    \label{fig:overK}
    \vspace{-0.4cm}
\end{figure}

Fig.~\ref{fig:overK} illustrates the minimum $P_k$ and $\gamma^\star$ as a function of $K$ for different $\kappa$ values. It is seen that increasing $K$ leads to higher $\gamma^\star$, which is expected due to the additional estimation complexity. Moreover, STC outperforms RAB up to a certain $K$ value, which depends on system parameters such as $P_t$ and $N_r$. For example, STC outperforms RAB and performs near-optimal up to $K = 4$ for this specific setup when $\kappa = 20$~dB. Interestingly, it is observed that decreasing $\kappa$ degrades the performance of STC and leads to RAB becoming favorable for lower $K$. Specifically, $\gamma^\star$ becomes larger for the channels with small $\kappa$. This is expected since the NLoS components of the channel become more significant, leading to a less accurate EB design due to only relying on the LoS components. Notably, the echo signal is also corrupted more significantly as the NLoS components become dominant.

\begin{figure}[t]
    \centering
    \centering
    \includegraphics[width=0.5\columnwidth]{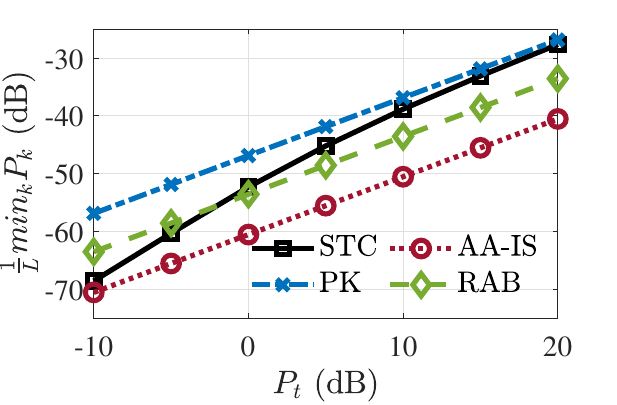} 
    \hspace{-5mm}
    \includegraphics[width=0.5\columnwidth]{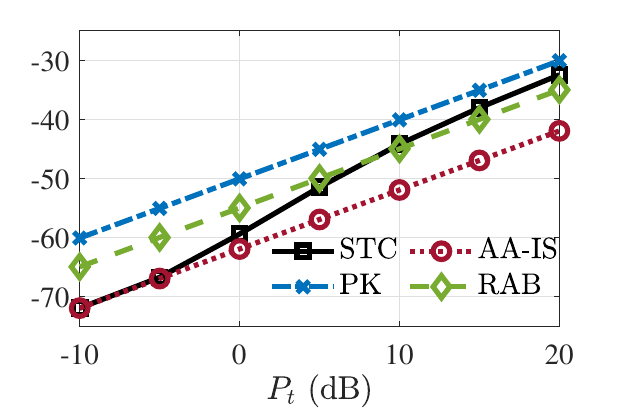} \\
    \vspace{-5.8mm}
    \includegraphics[width=0.5\columnwidth]{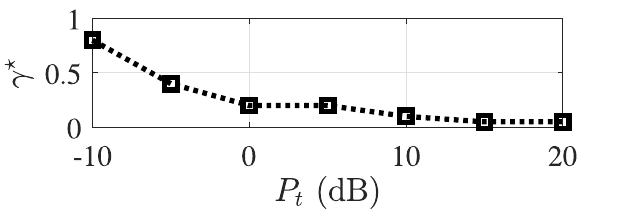} 
    \hspace{-5mm}
    \includegraphics[width=0.5\columnwidth]{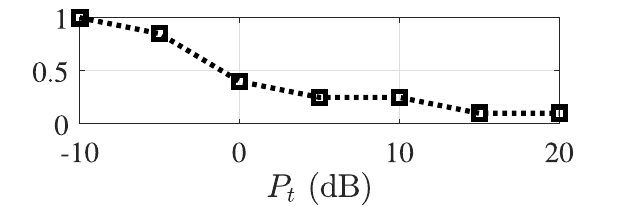}
    \caption{(a) Average minimum $P_k$ (top) and (b) $\gamma^\star$ (bottom) as a function of $P_t$ for $K = 2$ (left) and $K = 4$ (right). We assume $N_r = 24$ and $N_t = 12$.}
    \label{fig:overPt}
    \vspace{-0.4cm}
\end{figure}

Fig.~\ref{fig:overPt} brings further insights into the impact of the parameters on the performance of STC by illustrating the average minimum $P_k$ and $\gamma^\star$ as a function of $P_t$. It is seen that increasing $P_t$ enhances the performance of all the approaches. However, it has a more significant impact on STC since it leads to more accurate estimations in the sensing phase for lower $\gamma^\star$ by increasing the signal-to-noise ratio. For example, we observe that STC is outperformed by RAB when $P_t = -10$~dB. However, it outperforms RAB and approaches near-optimal performance with increasing $P_t$.

\begin{figure}[t]
    \centering
    \includegraphics[width=0.5\columnwidth]{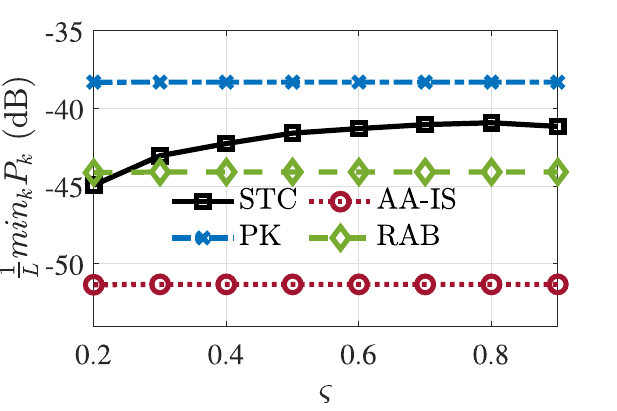} 
    \hspace{-5mm}
    \includegraphics[width=0.5\columnwidth]{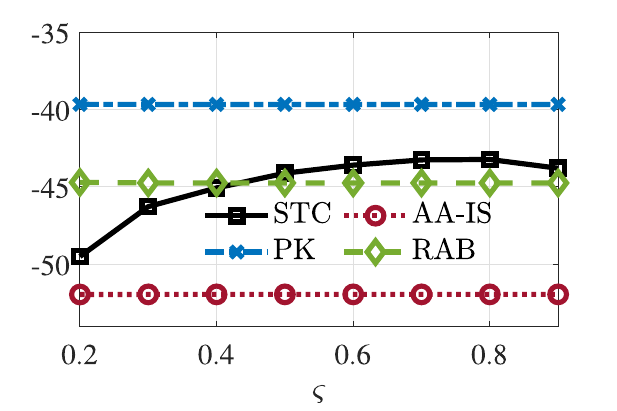} \\
    \vspace{-5.5mm}
    \includegraphics[width=0.5\columnwidth]{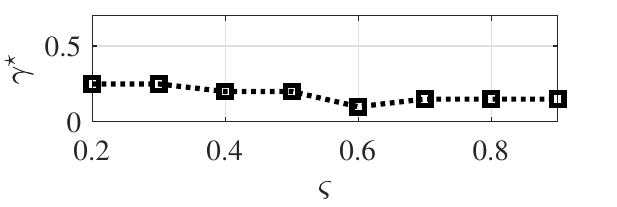} 
    \hspace{-5mm}
    \includegraphics[width=0.5\columnwidth]{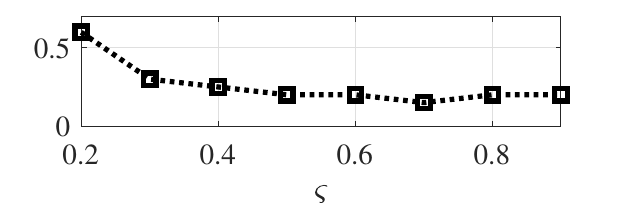}
    \caption{(a) Average minimum $P_k$ (top) and (b) $\gamma^\star$ (bottom) as a function of $\varsigma$ for $K = 3$ (left) and $K = 4$ (right). We assume $N = 40$ and $P_t = 10$~dB.}
    \label{fig:overantenna}
    \vspace{-0.4cm}
\end{figure}

For the analysis of the trade-off between $N_t$ and $N_r$, we introduce an additional parameter $\varsigma$ such that  $N_r = \lceil \varsigma N \rceil$ and $N_t = N - N_r$. Fig.~\ref{fig:overantenna} illustrates the average minimum $P_k$ and $\gamma^\star$ as a function of $\varsigma$. It is observed that increasing $\varsigma$ leads to improving the STC performance by providing sufficiently accurate estimations using a larger $N_r$. However, the performance enhances up to a specific $\varsigma$ value and after that, it starts to degrade because of the small $N_t$. For example, the STC performance improves up to $\varsigma = 0.8$ and outperforms RAB for $K = \{3,4\}$, but it starts to degrade afterward.


\section{Conclusions}\label{sec:conclude}

This paper considered a dual-functional WPT and sensing system where a BS charges multiple unresponsive devices with unknown locations over several transmission blocks. We formulated the charging problem to maximize the minimum received power at the devices and proposed the STC protocol, which divides the transmission blocks into sensing and WPT phases. First, the BS estimates the device locations by sending sensing signals and performing the MUSIC algorithm on the received echo. Then, the obtained information is used for WPT beamforming. The simulation results demonstrated that STC could outperform other benchmarks and achieve near-optimal performance with suitable transmit power, receive antennas, and sensing/charging time allocation, especially for fewer devices. We also highlighted the trade-off between the number of sensing and charging blocks and their impact on received power and estimation accuracy.



\bibliographystyle{ieeetr}
\bibliography{ref_abbv}

\end{document}